\documentclass[universe,article,accept,pdftex,moreauthors]{Definitions/mdpi}

\firstpage{1} 
\pubvolume{1}
\issuenum{1}
\articlenumber{0}
\pubyear{2024}
\copyrightyear{2024}

\datereceived{ } 
\daterevised{ } 
\dateaccepted{ } 
\datepublished{ } 

\hreflink{https://doi.org/} 

\usepackage{subcaption}
\usepackage{amsmath}
\usepackage{amssymb}

\usepackage{multirow}
\usepackage{longtable}

\Title{VLBI Analysis of a Potential High-energy Neutrino Emitter Blazar}

\TitleCitation{VLBI analysis of a potential high-energy neutrino emitter blazar}

\Author{Janka K\H{o}m\'{\i}ves $^{1}$ \orcidA{}, Krisztina \'Eva Gab\'anyi $^{* 1, 2, 3, 4}$ \orcidB{}, S\'andor Frey $^{3, 4, 5}$ \orcidC{} and Emma Kun $^{6, 7, 8, 3, 4}$ \orcidD{}}

\AuthorNames{Janka K\H{o}m\'{\i}ves, Krisztina \'Eva Gab\'anyi, S\'andor Frey and Emma Kun}

\AuthorCitation{K\H{o}m\'{\i}ves, J.; Gab\'anyi, K. \'E.; Frey, S.; Kun, E.}

\address{$^{1}$ \quad Department of Astronomy, Institute of Physics and Astronomy, ELTE E\"otv\"os Lor\'and University, P\'azm\'any P\'eter s\'et\'any 1/A, H-1117 Budapest, Hungary; k.gabanyi@astro.elte.hu (K\'EG); xd3y0o@student.elte.hu (JK)\\
$^{2}$ \quad HUN-REN--ELTE Extragalactic Astrophysics Research Group, ELTE E\"otv\"os Lor\'and University, P\'azm\'any P\'eter s\'et\'any 1/A, H-1117 Budapest, Hungary\\
$^{3}$ \quad Konkoly Observatory, HUN-REN Research Centre for Astronomy and Earth Sciences, Konkoly Thege Mikl\'{o}s \'{u}t 15-17, 1121 Budapest, Hungary; frey.sandor@csfk.org (SF)\\
$^{4}$ \quad CSFK, MTA Centre of Excellence, Konkoly Thege Mikl\'{o}s \'{u}t 15-17, 1121 Budapest, Hungary\\
$^{5}$ \quad Institute of Physics and Astronomy, ELTE E\"otv\"os Lor\'and University, P\'azm\'any P\'eter s\'et\'any 1/A, H-1117 Budapest, Hungary\\
$^{6}$ \quad Theoretical Physics IV, Faculty for Physics \& Astronomy, Ruhr University Bochum, 44780 Bochum, Germany; kun.emma@csfk.org (EK)\\
$^{7}$ \quad Faculty for Physics \& Astronomy, Astronomical Institute, Ruhr University Bochum, 44780 Bochum, Germany\\
$^{8}$ \quad Ruhr Astroparticle and Plasma Physics Center, Ruhr-Universit\"at Bochum, 44780 Bochum, Germany}

\abstract{Recent studies suggest that high-energy neutrinos can be produced in the jets of blazars, radio-loud active galactic nuclei (AGN) with jets pointing close to the line of sight. Due to the relatively poor angular resolution of current neutrino detectors, several sources can be regarded as the possible counterpart of a given neutrino event. Therefore, follow-up observations of counterpart candidates in the electromagnetic regime are essential. Since the Very Long Baseline Interferometry (VLBI) technique provides the highest angular resolution to study the radio jets of blazars, a growing number of investigations are conducted to connect individual blazars to given high-energy neutrino events. We analysed more than 20 years of available archival VLBI data of the blazar CTD\,74, which has been listed as a possible counterpart of a neutrino event. Using cm-wavelength data, we investigated the jet structure, determined the apparent speed of jet components, and the core flux density before and after the neutrino event. Our results indicate stationary jet features and a significant brightening of the core after the neutrino event.}


\begin{document}

\section{Introduction}

Blazars belong to the group of radio-loud active galactic nuclei (AGN) with their jets pointing in a small angle to the line of sight \cite{UrryPadovani}. In recent studies \cite{2020MNRAS.497..865G}, blazars were associated with high-energy neutrino events detected by the IceCube Neutrino Observatory located at the geographical South Pole \cite{Icecube}.  While as of now, except for the first blazar associated to a neutrino event, TXS\,0506$+$056 \cite{txs0i506}, no definitive link has been established between individual high-energy IceCube events and radio-loud AGN, there have been several assertions of associations with AGN at various statistical levels \cite{plavinsources, Plavin, Hovatta_2021, EmmaKun, 2023MNRAS.526..347N}. However, other authors found that the contribution of the blazar class as a whole to the IceCube neutrino signal is not dominant \cite{2023ApJ...955L..32B, Aartsen_blazar_neutrino}. Other studies using very similar blazar catalogs found no meaningful correlation between IceCube neutrino events and radio-bright AGN \cite{2021PhRvD.103l3018Z, 2023ApJ...954...75A}. In addition to the IceCube, similar studies were conducted with the ANTARES neutrino telescope \cite{2023arXiv230906874A}, which hinted a possible connection between the ANTARES-detected neutrino candidates with blazars.
Apart from blazars, other candidates for the source of high-energy neutrinos were proposed, such as gamma-ray bursts, supernova remnants, starburst galaxies \citep[e.g.,][]{neutrino_origin1}, tidal disruption events \cite{neutrino_tde}, white dwarf mergers \cite{neutrino_WD}, and other non-blazar AGN, e.g., NGC\,1068 \citep{NGC1068}. Thus, since the first association, analysing possible neutrino-source blazars has become an important task \cite{2023MNRAS.519.1396S, Acharyya, 2024MNRAS.527L..26S} as these objects may help us understand the astrophysical processes that lead to the production of high-energy neutrinos. 

In this paper, we present high-resolution Very Long Baseline Interferometry (VLBI) analysis of the blazar CTD\,74 (alternatively: TXS\,1123$+$264, J1125$+$2610) that has been statistically linked to an extremely high-energy track-like neutrino alert (EHEA2012-05-23) observed in $2012$, due to its positional proximity. CTD\,74 was in the sample of radio-loud AGN having VLBI observations at $8$\,GHz, and monitored at $22$\,GHz by the Russian RATAN-600 radio telescope \cite{1979S&T....57..324K} that were analyzed in connection with high-energy ($>200$~TeV) track-like neutrino events detected by the IceCube instrument \cite{Plavin}. The selection criteria of Extremely High Energy Alerts and alert-like events are described in \cite{Aartsen_ehea}. The directions of such events can be determined with an uncertainty of less than $1^{\circ}$ \cite{aartsen_2014a}, and CTD\,74 fell into the error region of such event. The positional association of the high-energy event combined with the available $25$-yr long multi-frequency VLBI observations, which covered the time of the neutrino event, made this object a promising candidate to study individually in the context of possible neutrino-emitter blazars. However, we note that the consideration of this particular blazar as a high-energy neutrino source was based on positional coincidence in previous works. We emphasize the possibility that the detected neutrino event might originate from a different source or might not even be astrophysical. Individual AGN sources have been studied in the past in a similar manner \cite{Kun_2023, Eppel, Kovalev_VLBIsource, NGC1068}. However, we are not aware of any study concerning the high-resolution radio emission of CTD\,74.

CTD\,74 \cite{Roma} is a blazar at a redshift of $z=2.3502 \pm 0.0001$ \cite{SDSS}. Its coordinates in the International Celestial Reference Frame (ICRF) are $11^{\mathrm{h}} 25^{\mathrm{m}} 53.7119^{\mathrm{s}}$ right ascension and $+26^{\circ} 10^{\prime} 19.979^{\prime\prime}$ declination \cite{ICRF}.
The source has been observed in several epochs with VLBI, but to our knowledge, a comprehensive analysis of the data has not been published yet. 
It is not included in the latest \textit{Fermi} Large Area Telescope catalog as a $\gamma$-ray source \cite{Abdollahi_2020} and no information was found about its X-ray detection.

Assuming a flat $\Lambda$ Cold Dark Matter cosmological model with Hubble constant $H_0=70$~km\,s$^{-1}$\,Mpc$^{-1}$, matter density parameter $\Omega_\mathrm{m}= 0.27$, and vacuum energy density parameter $\Omega_\mathrm{vac} = 0.73$, the angular diameter distance of the source is $D_\mathrm{A}= 1734.2$\,Mpc, and the angular scale is $8.407$\,pc\,mas$^{-1}$ \cite{wright}.

\section{Observations and data reduction }

We analyzed multi-frequency, multi-epoch pre-calibrated VLBI visibility data,  obtained from the {Astrogeo} (accessed on 28 September 2023) (\url{http://astrogeo.org}, {see also} \url{https://astrogeo.smce.nasa.gov/vlbi_images/}) website maintained by L. Petrov. Most of the observations of CTD\,74 were conducted at the $8$-GHz frequency band (central frequencies $7.62-8.67$\,GHz) in $23$ epochs between 1996 and 2021. The majority of these were astrometric observations conducted with a dual-band receiver, therefore in $19$ epochs, $2$-GHz-band measurements are also available. There were additional $3$ observations done at the $5$-GHz band (Table~\ref{tab:obs}). Since the central frequencies of the observations were slightly different from epoch to epoch, we refer to the three main bands of the observing frequencies as $8$, $5$, and $2$\,GHz in the text, while for calculations, we naturally use the central frequencies of the given data sets.

These observations were performed by heterogeneous arrays of radio telescopes. Primarily the Very Long Baseline Array (VLBA) of the U.S. National Radio Astronomy Observatory (NRAO) was used. It consists of $10$ antennas at Brewster (BR), Fort Davis (FD), Hancock (HN), Kitt Peak (KP), Los Alamos (LA), Mauna Kea (MK), North Liberty (NL), Owens Valley (OV), Pie Town (PT), and St. Croix (SC). But occasionally, multiple antennas from the European VLBI Network (EVN),  Medicina (MC, Italy), Sheshan (SH, China), Onsala (ON, Sweden), Hartebeesthoek 26-m (HH, South Africa), as well as other geodetic VLBI radio telescopes, Goddard (GG, USA), Gilmore Creek (GC, USA), Hartebeesthoek 15-m (HT, South Africa), Kokee Park (KK, USA), Matera (MA, Italy), Wettzell (WZ, Germany), Tsukuba (TS, Japan), Ny-{\AA}lesund (NY, Norway), Westford (WF, USA), and Yarragadee (YG, Australia) participated in various experiments. The details of the observations are summarized in Table \ref{tab:obs}. It is indicated with the corresponding two-letter station codes preceded by a minus sign if one or two antennas from the VLBA were missing from the array. 

\begin{figure}[H]
\centering
     \begin{subfigure}[t]{0.45\textwidth}
         \centering
         \includegraphics[width=\textwidth]{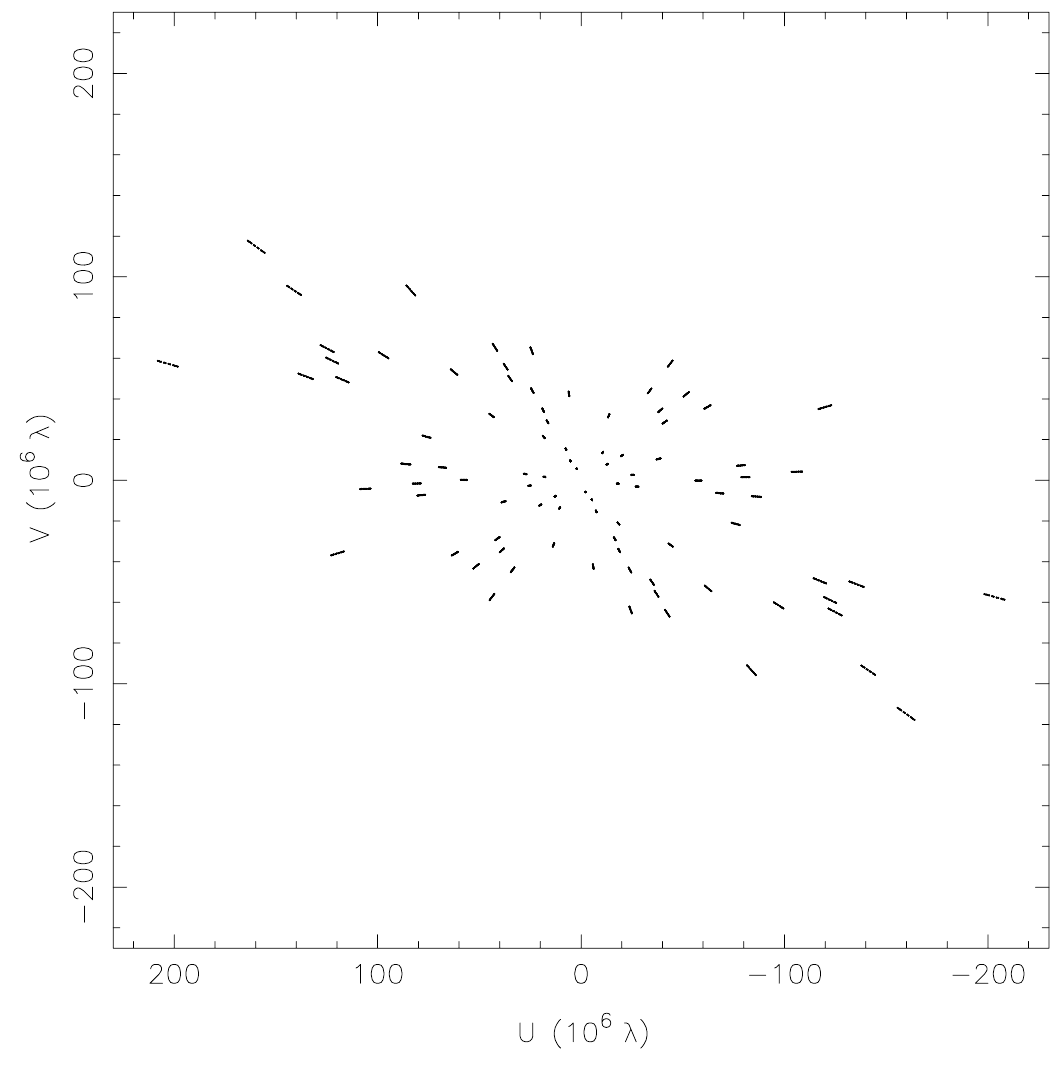}
     \end{subfigure}
     \hfill
     \begin{subfigure}[t]{0.45\textwidth}
         \centering
         \includegraphics[width=\textwidth]{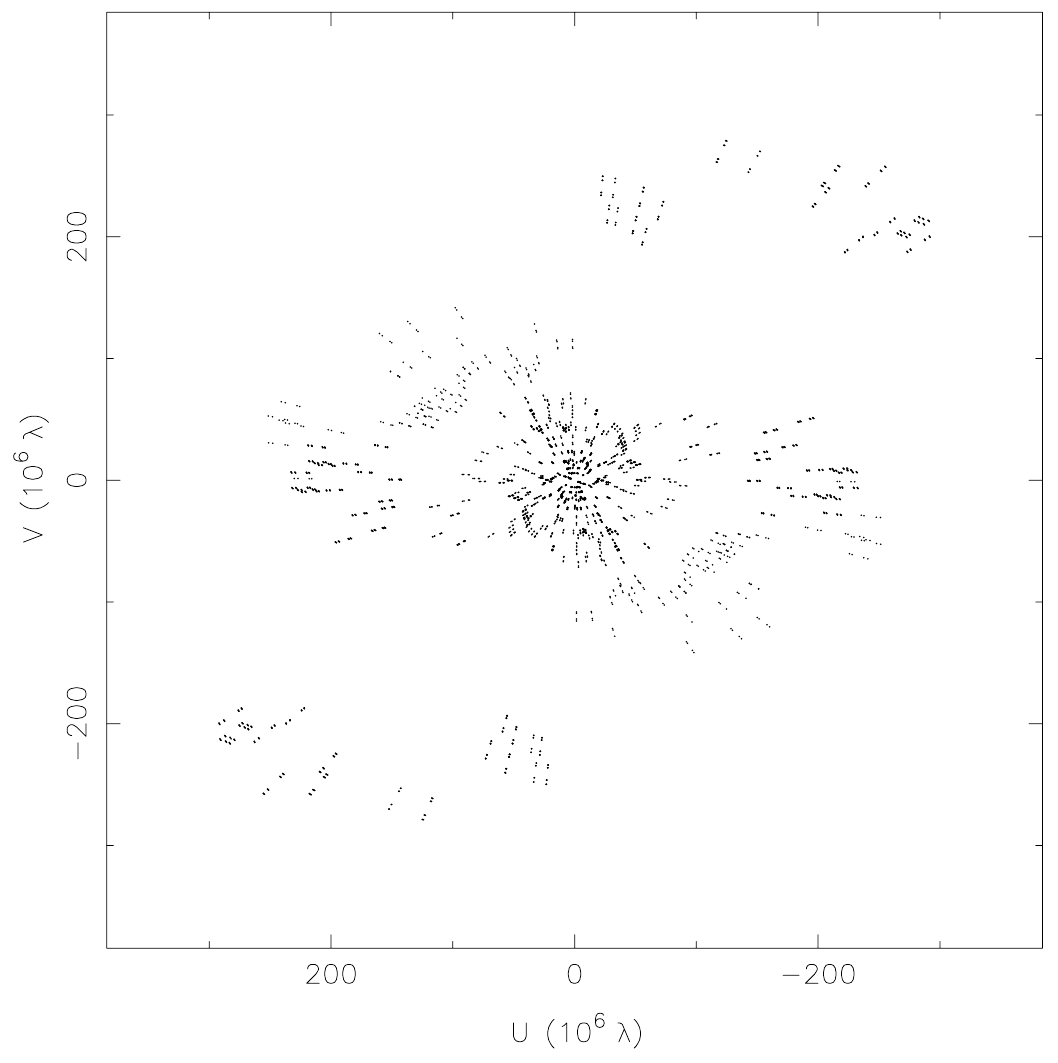}
    \end{subfigure}
    \caption{The $(u,v)$ coverages of two observations performed at $8.6$\,GHz. The axes represent the baseline vectors in $u$ and $v$ coordinates in units of million wavelengths. {\it Left:} The $(u,v)$ coverage of the observation performed on 2017 March 27 with VLBA-only baselines. {\it Right:} The $(u,v)$ coverage of the observation performed on 2020 November 18 with VLBA and global baselines.}
    \label{fig:uvplot}
\end{figure}

It is important to note that the data set is not homogeneous, with vastly different on-source integration times (as seen in Table~\ref{tab:obs}) resulting in different sampling of the visibility functions that can affect the quality of the obtained images. In Figure \ref{fig:uvplot}, we show the distributions of baseline vectors, commonly referred as $(u,v)$ coverage, for two observations, one with the shortest integration time ($14$\,s, performed on 2017 March 27), and one with a significantly longer integration time ($12\,892$\,s, performed on 2020 November 18).

\begin{table}
\centering
\caption{Details of the analysed VLBI observations. The observational epochs, central frequencies in GHz ($\nu$), the codes of the participating antennas (Stations), the on-source integration times of the measurements, the number of intermediate frequency channels (IFs), their bandwidth in MHz (BW), and the project codes and references (where applicable) are given in the columns. \label{tab:obs}}
	\setlength\tabcolsep{2pt}
	\newcolumntype{C}{>{\centering\arraybackslash}X}
	\begin{adjustwidth}{-\extralength}{0cm}
	 \begin{tabularx}{\fulllength}{C C C C C C} 
	\toprule
            \toprule
	\textbf{Epoch}	& \textbf{$\nu$ (GHz)}	& \textbf{Stations} & \textbf{On-source time (s) } & \textbf{IF $\times$ BW [MHz]} & \textbf{Ref.} \\
			\midrule		
\multirow{2}{*}{1996.05.15.}  & $8.34$  & \multirow{2}{*}{VLBA}  & \multirow{2}{*}{$282$} & \multirow{2}{*}{$4 \times 8$} & \multirow{2}{*}{BB023 \cite{vcs1}} \\ & $2.27$ & \\ 
\midrule
1996.06.06. & $4.92$ & VLBA & $140$ & $8 \times 8$ & BH019 \cite{prelaunch}\\ \midrule
\multirow{2}{*}{1997.01.10.} & $8.34$ & \multirow{2}{*}{VLBA} & \multirow{2}{*}{$556$} & \multirow{2}{*}{$4 \times 8$} & \multirow{2}{*}{BF025\cite{pushkarev2012}} \\ & $2.29$ &\\ \midrule
\multirow{2}{*}{2000.05.22.} & $8.65$ & \multirow{2}{*}{\shortstack[t]{VLBA, GC, GG, HH, KK, \\ MA, MC, NY, TS, WF, WZ}} & \multirow{2}{*}{1352} & \multirow{2}{*}{$4 \times 8$} & \multirow{2}{*}{RDV21 \cite{2008evn}} \\ & $2.29$ &\\ \midrule
\multirow{2}{*}{2002.09.25.} & $8.65$ & \multirow{2}{*}{\shortstack[t]{VLBA, GC, GG, HH, KK, \\ MA, MC, NY, TS, WF, WZ}} & \multirow{2}{*}{$1548$} & \multirow{2}{*}{$4 \times 8$} & \multirow{2}{*}{RDV35 \cite{pushkarev2012}}\\ & $2.30$ & \\ \midrule
2006.02.09. & $4.85$ & VLBA & $12280$ & $4 \times 8$ & BT085 \cite{vips}\\ \midrule
2011.02.27. & $8.36$ & VLBA, HH & $280$ & $8 \times 16$ & BC196 \cite{vlbi_2mass} \\ \midrule
2011.08.06. & $8.36$ & VLBA & $383$ & $8 \times 16$ & BC196 \cite{vlbi_2mass} \\ \midrule
2012.04.10. & $8.36$ & VLBA & $96$ & $8 \times 16$ & BC201 \cite{vlbi_2mass} \\ 
\midrule
\multirow{2}{*}{2013.07.24.} & $8.64$ & \multirow{2}{*}{\shortstack[t]{HH, MA, NY, WF, \\ VLBA $-$BR, $-$FD}} & \multirow{2}{*}{$17383$} & \multirow{2}{*}{$4 \times 8$} & \multirow{2}{*}{RV100} \\ & $2.30$ \\ \midrule
\multirow{2}{*}{2014.02.12.} & $8.64$ & \multirow{2}{*}{\shortstack[t]{VLBA, HT, MA, \\ NY, ON, WZ}} & \multirow{2}{*}{$7259$} & \multirow{2}{*}{$4 \times 8$} & \multirow{2}{*}{RV103} \\ & $2.30$ \\ \midrule
\multirow{2}{*}{2014.05.31.} & $8.67$ & \multirow{2}{*}{VLBA} & \multirow{2}{*}{$609$} & $12 \times 32$ & \multirow{2}{*}{BG219 \cite{vcs2}} \\ & $2.29$ & & &  $4 \times 32$\\ \midrule
\multirow{2}{*}{2015.01.23.} & $8.67$ & \multirow{2}{*}{VLBA} & \multirow{2}{*}{25} & $12 \times 32$ &\multirow{2}{*}{BG219 \cite{vcs2}} \\ & $2.29$ & & &  $4 \times 32$ \\ \midrule
\multirow{2}{*}{2016.07.17.} & $7.62$ & \multirow{2}{*}{VLBA} & \multirow{2}{*}{$56$} & \multirow{2}{*}{$8 \times 32$} & \multirow{2}{*}{BG192 \cite{vcs2}}\\ & $4.34$ \\ \midrule
\multirow{2}{*}{2016.11.30.} & $8.65$ & \multirow{2}{*}{VLBA, WZ, HH, ON} & \multirow{2}{*}{$31760$} & \multirow{2}{*}{$4 \times 8$} & \multirow{2}{*}{RV120}\\ & $2.29$ \\ \midrule
\multirow{2}{*}{2017.03.27.} & $8.67$ & \multirow{2}{*}{VLBA} & \multirow{2}{*}{$14$} & $12 \times 32$ & \multirow{2}{*}{UF001 \cite{hunt}} \\ & $2.29$ & & &  $4 \times 32$ \\ \midrule
\multirow{2}{*}{2017.04.28.} & $8.67$ & \multirow{2}{*}{VLBA} & \multirow{2}{*}{$37$} & $12 \times 32$ & \multirow{2}{*}{UF001 \cite{hunt}} \\ & $2.29$ & & &  $4 \times 32$\\ \midrule
\multirow{2}{*}{2017.06.28.} & $8.67$ & \multirow{2}{*}{\shortstack[t]{VLBA, HH, MC, NY, ON, \\ YG, WN, WZ}} & \multirow{2}{*}{$17352$}  & \multirow{2}{*}{$4 \times 8$} & \multirow{2}{*}{RV123} \\ & $2.29$ \\ \midrule
\multirow{2}{*}{2017.09.26.} & $8.67$ & \multirow{2}{*}{VLBA $-$SC} & \multirow{2}{*}{$47$} & $12 \times 32$ & \multirow{2}{*}{UF001 \cite{hunt}} \\ & $2.28$ & & &  $4 \times 32$\\ \midrule
\multirow{2}{*}{2017.10.09.} & $8.67$ & \multirow{2}{*}{VLBA $-$SC} & \multirow{2}{*}{$35$} & $12 \times 32$ & \multirow{2}{*}{UF001 \cite{hunt}} \\ & $2.29$ & & &  $4 \times 32$ \\ \midrule
\multirow{2}{*}{2018.07.05.} & $8.65$ & \multirow{2}{*}{VLBA $-$NL} & \multirow{2}{*}{$119$} & $12 \times 32$ & \multirow{2}{*}{UG001} \\ & $2.29$ & & &  $3 \times 8$ \\ \midrule
\multirow{2}{*}{2020.11.18.} & $8.64$ & \multirow{2}{*}{HH, ON, WZ, VLBA $-$SC} & \multirow{2}{*}{$12892$} & \multirow{2}{*}{$4 \times 16$} & \multirow{2}{*}{RV144}\\ & $2.28$ \\ \midrule
\multirow{2}{*}{2021.03.24.} & $8.64$ & \multirow{2}{*}{ON, SH, WZ, VLBA $-$MK} & \multirow{2}{*}{$20238$} & \multirow{2}{*}{$4 \times 16$} & \multirow{2}{*}{RV146} \\ & $2.28$ \\ \midrule
\multirow{2}{*}{2021.05.19.} & $8.64$ & \multirow{2}{*}{\shortstack[t]{VLBA, HH, NS, NY, ON, \\ SH, WZ}} & \multirow{2}{*}{$16018$} & \multirow{2}{*}{$4 \times 16$} & \multirow{2}{*}{RV147} \\ & $2.28$ \\ \midrule
\multirow{2}{*}{2021.07.07.} & $8.64$ & \multirow{2}{*}{NY, VLBA $-$NL} & \multirow{2}{*}{$4446$} & \multirow{2}{*}{$4 \times 16$} & \multirow{2}{*}{RV148}\\ & $2.28$ \\ 
\bottomrule
\bottomrule
 \end{tabularx}
 \end{adjustwidth}

\end{table}

\subsection{Hybrid mapping and brightness distribution modeling}
\label{hybrid mapping}

\begin{figure}[H]
\centering
     \begin{subfigure}[t]{0.45\textwidth}
         \centering
         \includegraphics[width=\textwidth]{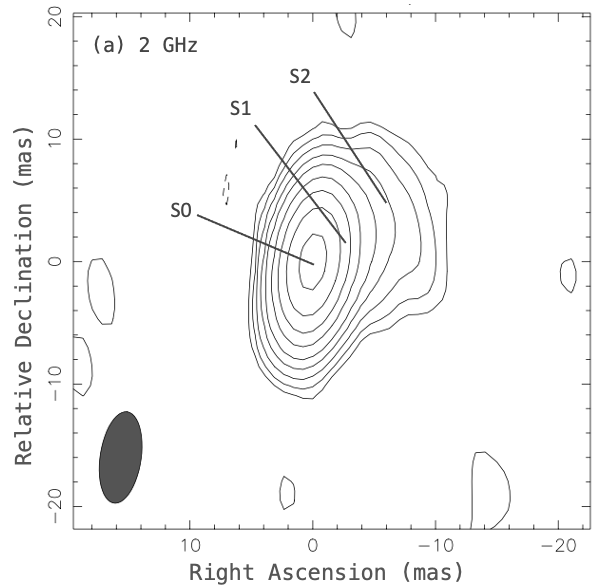}
         \label{fig:clean_map_S}
         \end{subfigure}
     \hfill
     \begin{subfigure}[t]{0.45\textwidth}
         \includegraphics[width=\textwidth]{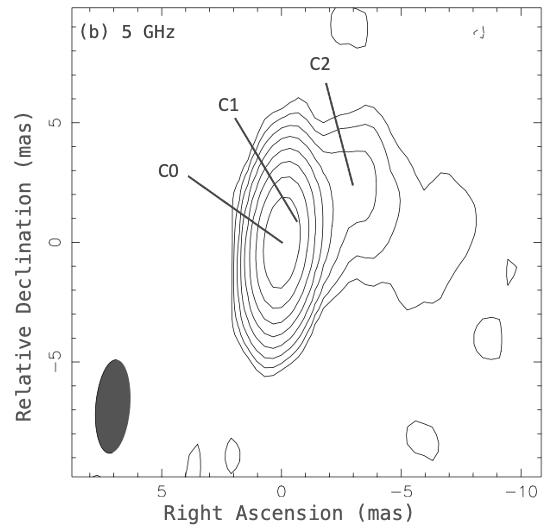}
         \label{fig:clean_map_C}
     \end{subfigure}
     \begin{subfigure}[b]{\textwidth}
         \centering
         \includegraphics[width=0.45\textwidth]{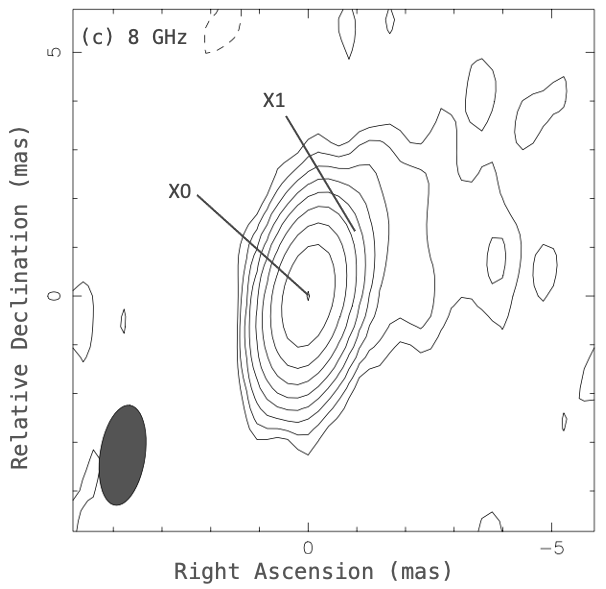}
         \label{fig:clean_map_X}
     \end{subfigure}
     \caption{VLBI images of CTD\,74 obtained through hybrid mapping for the measurements taken in 1996 at $2$, $5$, and $8$\,GHz in panels (a), (b), and (c), respectively. In all maps, the positions of the Gaussian components fitted to the visibility data are marked. {\it Panel (a):} The peak intensity is $1160$ mJy\,beam$^{-1}$. The lowest contours are at $\pm 3.44$\,mJy\,beam$^{-1}$. The restoring beam size is $7.54\mathrm{\,mas} \times 3.34\mathrm{\,mas}$. The position angle of the major axis is $\mathrm{PA}=-8.35^{\circ}$. {\it Panel (b):} The peak intensity is $1020$\,mJy\,beam$^{-1}$. The lowest contours are at $\pm 5.44$\,mJy\,beam$^{-1}$. The restoring beam size is $3.92\mathrm{\,mas} \times 1.44\mathrm{\,mas}$, $\mathrm{PA} = -4.63 ^{\circ}$. {\it Panel (c):} The peak intensity is $746$\,mJy\,beam$^{-1}$. The lowest contours are at $\pm 4.45$\,mJy\,beam$^{-1}$. The restoring beam size is $2.07 \mathrm{\,mas}\times 0.926\mathrm{\,mas}$, $\mathrm{PA} = -8.59 ^{\circ}$. In each map, the lowest positive contour is drawn at $3\sigma$ image noise level, and subsequent contours increase by a factor of two. The restoring beams are shown in the lower left corners of each panel.}
     \label{fig:clean_maps}
\end{figure}

We used the {Difmap software (version number 2.5k)} \cite{difmap} to produce images of CTD\,74 using the hybrid mapping method \cite[e.g.][]{Walker,2009NewAR..53..307F}. First, we created the dirty map via Fourier-transformation of the visibilities, after setting the appropriate map and pixel size. We defined a small region around the brightest pixel value in the dirty map, and using the the CLEAN algorithm \cite{Hogbom}, we subtracted point-source responses with $5\%$ of the brightest pixel value from the image in $50$ or $100$ iterations. This first CLEAN component model was then used to calibrate the phases of the visibilities in the phase self-calibration process \cite{Cornwell}. Then a new, residual map, from which the previous CLEAN component model had already been subtracted, was checked for the value and position of the brightest pixel, and the above described procedure was repeated. In order not to include spurious features in the point source model, i.e. to avoid cleaning the artifacts arising in the residual maps from the non-ideal visibility sampling, the brightest pixel search was always guided by the definitions of small regions for the cleaning process. That way the point source model was iteratively refined in small steps, in parallel with the improvement of the phase self-calibration.

When the model could not be improved further as judged from the noise level in the residual map, typically when the brightest pixel in the residual image was $\lesssim 5$ times of the noise level, we performed iterations of amplitude and phase self-calibration, gradually decreasing the solution interval by a factor of two in each iteration starting with a few hours down to $1-2$ minutes. We note here that because the noise in the residual image does not follow Gaussian statistics, the usage of root mean square value to estimate the total uncertainty is not strictly correct \cite{hovatta_rms, Gabuzda_rms}. The maps shown in Figure~\ref{fig:clean_maps} are chosen as examples to represent the final images that were obtained for all epochs and frequencies. 

In order to quantify the source brightness distribution, we fitted the self-calibrated visibility data with two-dimensional Gaussian model components \cite{1995ASPC...82..267P}. The parameters of these components can be found in Tables~\ref{8ghz}, \ref{2ghz}, and \ref{5ghz} for the $8$-, $2$-, and $5$-GHz data, respectively. To minimize the number of free parameters, we initially attempted to fit circular Gaussian components only. All four parameters of the components, positions, flux density ($S$ [Jy]), and full width at half maximum (FWHM) size ($W_1$ [mas]), were set to freely variable. In seven cases, we could not obtain a stable fit with using only circular Gaussian components, therefore we included elliptical components also. For those, the additional two parameters, the minor axis size ($W_2$ [mas]) and the position angle of the major axis ($\Phi$ [$^\circ$]) were free parameters as well. In most of these cases ($5$), the elliptical components were needed to describe the brightness distribution of the core, while in two other cases, a jet feature was fitted by elliptical Gaussians, as seen in Tables~\ref{8ghz}, \ref{2ghz}, and \ref{5ghz}.

\subsection{Error calculation of model parameters}

When calculating the errors of the model parameters, we used the formulae given in \cite{Emma2014, Schinzel}. These are based on the prescriptions of \cite{Image_Analysis}, however, they take into account the image artefacts arising from the sparse $(u,v)$ coverage of VLBI observations via the inclusion of the restoring beam.

We determined the post-fit root mean square error of the image, $\sigma_\mathrm{S_p}$ within a rectangular region around the position of the fitted components in the residual map.
The error of the component diameter (FWHM) is given as: 

\begin{linenomath}
\begin{equation}
    \begin{aligned}
    \sigma_{\mathrm{W}_i} & = \begin{cases} \sigma_{\mathrm{S}_\mathrm{p}} \sqrt{\theta^2 + W_i^2} / S_\mathrm{p}\text{,}  \hspace{15pt} \theta > W_i \\
     \sigma_\mathrm{S_\mathrm{p}} W_i / S_\mathrm{p}\text{,}  \hspace{50pt} \theta \leq W_i \text{,}
    \end{cases}
    \end{aligned}
\end{equation}
\end{linenomath}
where $S_\mathrm{p}$ is the peak intensity of the components with an error of $\sigma_\mathrm{S_p}$. 
$\theta$ denotes the size of the restoring beam as $\theta = \sqrt{\theta^2_\mathrm{min} + \theta^2_\mathrm{maj}}$, where $\theta_\mathrm{min}$ is the minor and $\theta_\mathrm{maj}$ is the major axis of the restoring beam. The index $i$ can take a value of $1$ and $2$ for the two axes of the fitted component. 

If the size of a component is smaller than the smallest resolvable size with the interferometer array of the given observation, only an upper limit of the source size, $\theta_\textrm{lim}$, can be given. To calculate $\theta_\textrm{lim}$, we used the equation in \cite{Blasars}:

\begin{linenomath}
\begin{equation}
    \theta_{\textrm{lim}} =\theta_\mathrm{maj} \sqrt{\frac{4 \ln{2}}{\pi} \ln{\left(\frac{\textrm{SNR}}{\textrm{SNR}-1}\right)}} \text{,}
\end{equation}
\end{linenomath}
where SNR stands for the signal-to-noise ratio, calculated from the ratio of peak intensity to the noise level.

The error of the flux density ($\sigma_\mathrm{S}$) includes the squared sum of the error given by the analytical formula in \cite{Image_Analysis}, and the contribution arising from the amplitude calibration error:
\begin{linenomath}
\begin{equation}
     \sigma_\mathrm{S} = \sqrt{\sigma_\mathrm{S_p}^2 + \left(S \cdot \sigma_\mathrm{W} / W\right)^2+(0.05\cdot S)^2}
\end{equation}
\end{linenomath}
where $W = \sqrt{(W_1 + W_2)^2}$. The amplitude calibration error is commonly taken as $5$\,\% of the flux density in case of VLBA observations \cite{Homan2002}.

\subsection{Analysis}

\begin{figure}[H]
\includegraphics[width=12.5 cm]{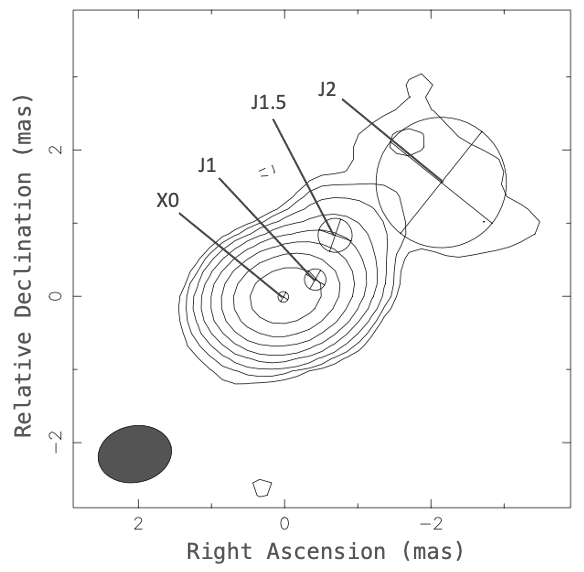}
\caption{VLBI image of CTD\,74 taken on 2014 Feb 12 at $8$\,GHz, created via Gaussian model fitting to the self-calibrated visibilities. The location and size (FWHM) of the model components are indicated with crosses and circles. The peak intensity is $867$\,mJy\,beam$^{-1}$. The lowest contour is at $\pm 3.7$\,mJy\,beam$^{-1}$, at 3 times the image noise level. The restoring beam size is $1.0\mathrm{\,mas}\times0.77\mathrm{\,mas}$, $\mathrm{PA} = -78.8^{\circ}$. It is shown in the lower left corner of the image.}
\label{fig:modelfit}
\end{figure}   

The source brightness distribution could be fitted with a core X0 and one or more additional jet components. For most of the $23$ epochs of $8$-GHz measurements, we could consistently identify two jet components, J1 and J2 (in $20$ epochs). In $7$ epochs, due to the denser sampling of the visibility function and the better sensitivity, we could fit an additional component (J1.5) located in between J1 and J2. An image from one of these measurements, the one performed on 2014 Feb 12, is shown in Figure~\ref{fig:modelfit} with the model components labeled. Finally, in the earliest two epochs, we could not securely identify the fitted jet components with those after 2000. These components are named as X1 and X2 (Table~\ref{8ghz}).

At $2$\,GHz, $11$ out of the $19$ epochs could be described with three Gaussian components. In the other cases, two Gaussian components were needed to adequately fit the visibility data (Table~\ref{2ghz}). The parameters of the fitted Gaussian model components at $5$~GHz in $3$ epochs are given in Table~\ref{5ghz}.

\section{Results}

In all the analysed data, CTD\,74 consistently showed a core--jet structure, with a short jet oriented to the northwest. Even at $2$\,GHz, the jet emission cannot be traced farther than $\sim10$\,mas from the core. This can be caused by the non-ideal sampling of the visibilities at the shortest baselines causing the observations to be less sensitive to the largest spatial scales of the emitting regions.

\subsection{Jet morphology in various epochs}

\begin{figure}[h]
\includegraphics[width=12.5 cm]{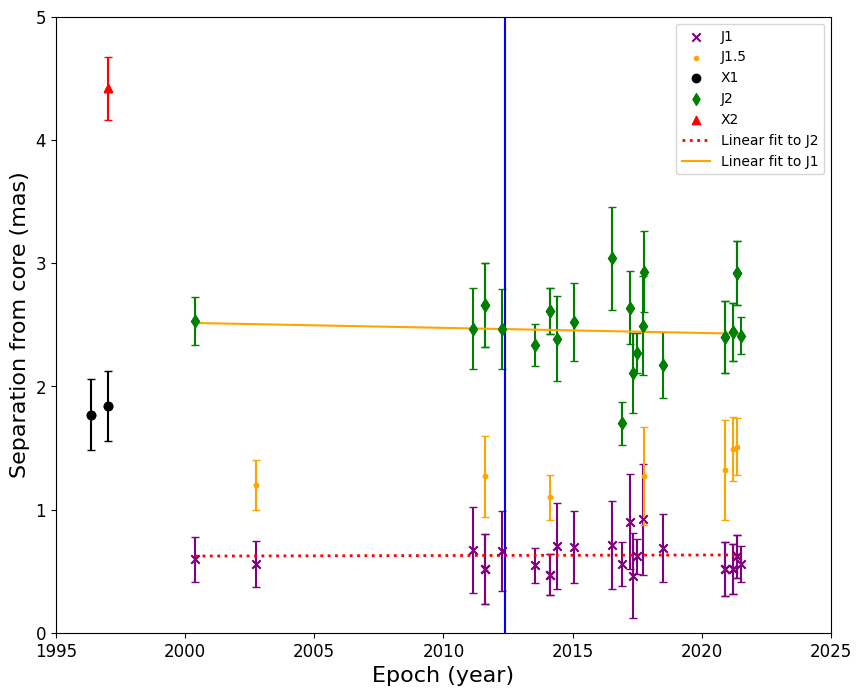}
\caption{Separation of the jet components from the core as a function of time at $8$\,GHz. The apparent proper motion of J1 and J2 components that could consistently be identified through most of the epochs was modeled with a linear function. These functions are indicated by the dashed red and solid yellow lines, respectively. The blue vertical line marks the time of the neutrino event EHEA2012-05-23 (2012 May 23). \label{fig:X_sep}}
\end{figure}   

\begin{figure}[H]
\includegraphics[width=12.5 cm]{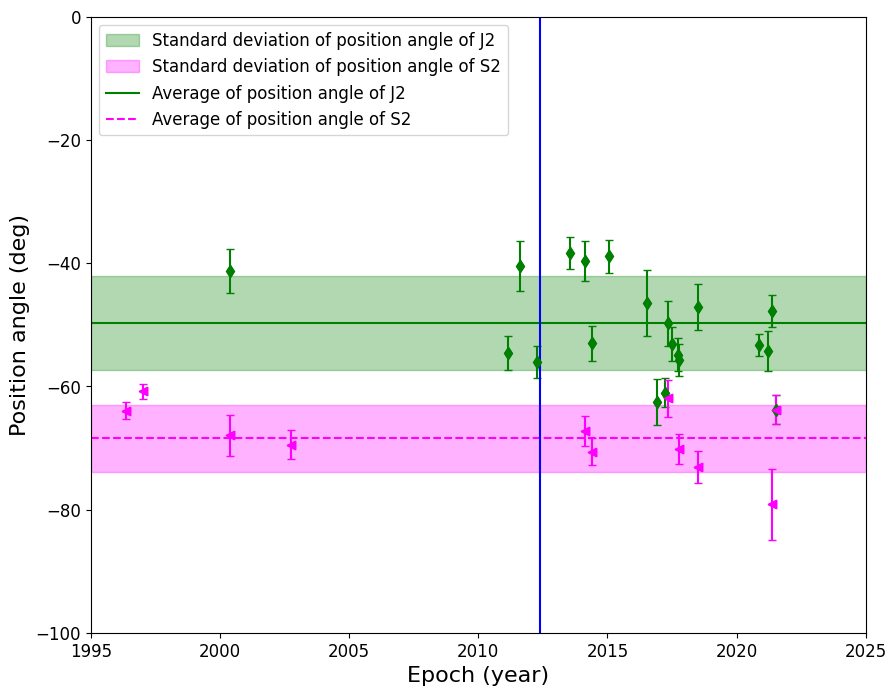}
\caption{The position angles of components J2 (green) and S2 (magenta) as a function of time. The angles are measured from north through east. The average and standard deviation of the position angles are indicated. The blue vertical line marks the time of the neutrino event EHEA2012-05-23 (2012 May 23). \label{fig:pos_angle}}
\end{figure}   

To study the apparent motion of the jet components, we calculated their separation from the core at each epoch. 
Since $8$\,GHz is the highest frequency and thus these observations provide the best angular resolution, the $8$-GHz data are the most suitable to detect structural changes and jet component motions in the compact environment around the core. The core separations of these jet components are shown in Figure~\ref{fig:X_sep}. 
We fitted linear functions to the core--jet component separations versus time, to determine the apparent proper motion of the jet components. The slopes of the lines fitted for the components J1 and J2 are $(-0.002 \pm 0.006)$\,mas\,yr$^{-1}$ and $(-0.009 \pm 0.008)$\,mas\,yr$^{-1}$, respectively. These values are consistent with no proper motion within the uncertainties (Figure~\ref{fig:X_sep}). Thus, no ballistic movement could be detected at $8$\,GHz in either of the jet components. We reached the same conclusion for the $2$-GHz data, after fitting linear functions to the core--jet components separations.

When studying core separation, we assume that the components are moving linearly, neglecting their two-dimensional motion in the plane of the sky. To investigate this motion, we examined how the position angles of the jet components change over time. In Figure~\ref{fig:pos_angle}, we show the position angles of the jet components that are the most distant from the core at the given frequency, and having a large number of detections, i.e., J2 at $8$\,GHz and S2 at $2$\,GHz. The average position angles are $(-50 \pm 8)^{\circ}$ and $(-69 \pm 6)^{\circ}$ for J2 and S2, respectively. In case of the components closer to the core, the average position angles are $(-52 \pm 8)^{\circ}$ and $(-71 \pm 7)^{\circ}$ for J1 and S1, respectively. Thus, the northwestern direction of the jet remains constant through the epochs, but there is a slight bend of $\sim 20^\circ$ between the inner $\lesssim 3$\,mas and outer regions of the jet. This could be traced with the observations taken at different frequencies.

In summary, the jet components appear stationary in the plane of the sky, not showing significant changes in the their core separation and position angle. 

\subsection{Flux density}

\begin{figure}[H]
\centering
     \begin{subfigure}[t]{0.9\textwidth}
         \centering
         \includegraphics[width=\textwidth]{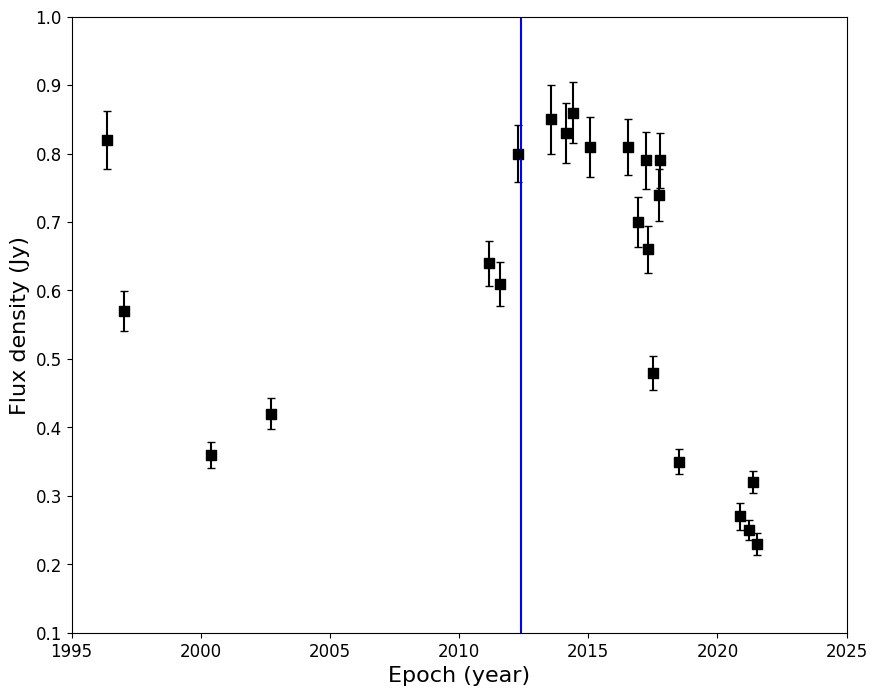}
         \end{subfigure}
     \hfill
     \begin{subfigure}[t]{0.9\textwidth}
         \includegraphics[width=\textwidth]{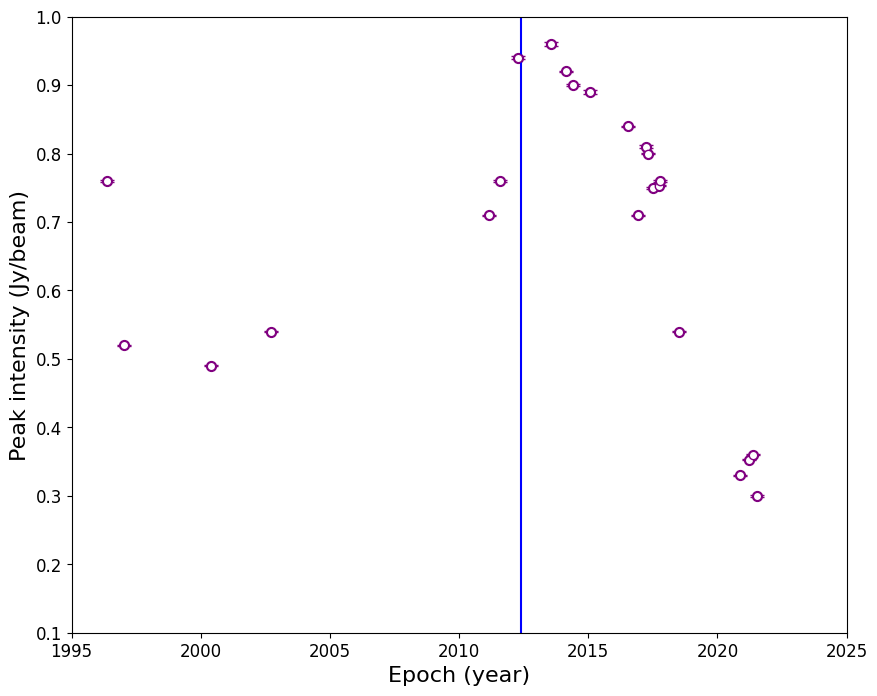}
     \end{subfigure}
     \caption{\textit{Top:} Flux density of the core component as a function of time at $8$\,GHz (black squares). \textit{Bottom:} Peak intensity (purple circles) in each image restored with the same beam (see details in the text). The errors of the peak intensities are comparable to the size of the symbol. The blue vertical line marks the time of the neutrino event EHEA2012-05-23 (2012 May 23).}
     \label{fig:flux+int}
\end{figure}

We examined how the flux density of the core component changes over time. The average flux density value of the core is $0.62$\,Jy at $8$\,GHz, but it shows significant variability during the $\sim15$\,yr of analysed VLBI observations. Most notably, it shows a brightening around 2014--2015 (see Figure~\ref{fig:flux+int}). During its brightening in 2014--2015, the core was $38\%$ brighter than its average value, reaching ($0.86\pm0.04$)\,Jy on 2014 May 31. Then the core faded back quickly after 2017. As there was no $8$-GHz VLBI observation taken of the object between 2003 and 2010, the core flux density changes cannot be discussed in this time range.

Since the flux density derived from the Gaussian components fitted to the visibilities may in principle be influenced by the heterogeneous resolution of the different observations, we also compared the peak intensity value of the images created with the same restoring beams, $0.98 \mathrm{\,mas}\times2.47$\,mas with a position angle of $-22.51^{\circ}$. We chose the observation with the shortest on-source time (2017 March 27). The same trend as in the fitted core component flux densities can be seen in the time variability of the intensity values as well (Figure\,\ref{fig:flux+int}).

We also examined the change in core flux density at $2$\,GHz. Here, with less frequent time sampling, we also observe the brightening and fading that was seen at $8$\,GHz.

\subsection{Brightness temperature}

\begin{figure}[H]
\includegraphics[width=12.5 cm]{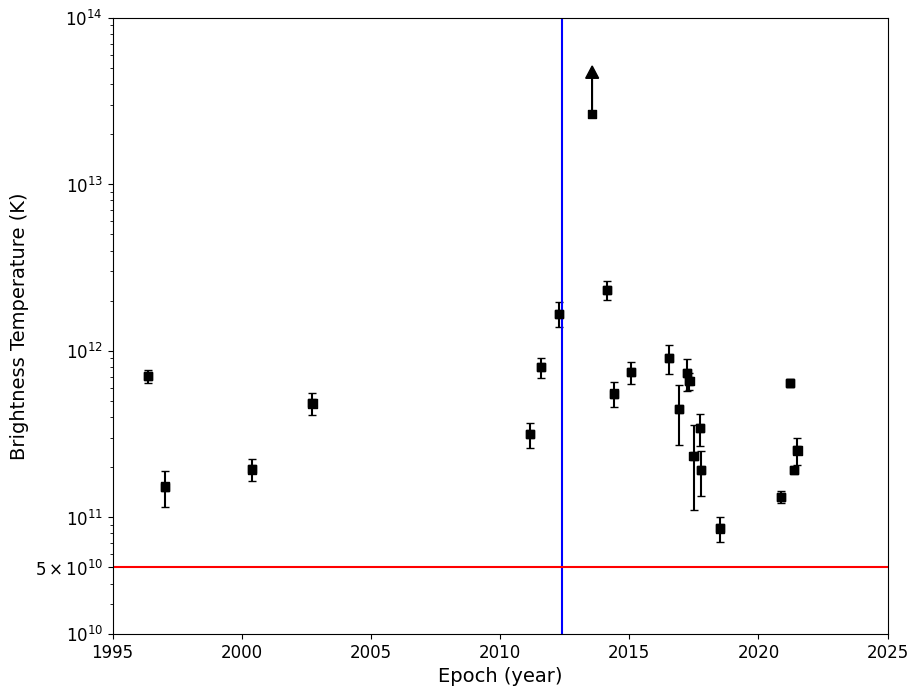}
\caption{Brightness temperature of the core component as a function of time, as measured at $8$\,GHz. The blue vertical line marks the time of the neutrino event EHEA2012-05-23 (2012 May 23). The red horizontal line indicates the value of the equipartition brightness temperature, $5\cdot10^{10}$\,K  \cite{Readhead}. \label{fig:X_TB}}
\end{figure}  

Knowing the redshift of the object, from the parameters of the fitted Gaussian components, their brightness temperatures, $T_\mathrm{b, VLBI}$, can be calculated \cite{Condon}. For the core component:

\begin{linenomath}
\begin{equation}
    T_{\mathrm{b,VLBI}}=1.22 \cdot 10^{12} (1 + z) \frac{S_\mathrm{core}}{\nu^2 W_1 W_2} \text{\, [K],}
\end{equation}
\end{linenomath}
where $S_\textrm{core}$ is the flux density expressed in Jy, $\nu$ the observing frequency in GHz. The component major and minor axes (FWHM), $W_1$ and $W_2$, are given in mas.

Assuming that the intrinsic brightness temperature of the source is equal to the equipartition value, $T_{\textrm{int}}=T_{\textrm{eq}} \approx 5\cdot10^{10}$\,K \cite{Readhead}, the Doppler boosting factor, which quantifies the relativistic beaming effect, can be calculated as follows:

\begin{linenomath}
\begin{equation}
    \delta = \frac{T_{\textrm{b,VLBI}}}{T_{\textrm{int}}} \approx \frac{T_{\textrm{b,VLBI}}}{T_{\textrm{eq}}} \text{.}
\end{equation}
\end{linenomath}

As seen in Figure~\ref{fig:X_TB}, every measured brightness temperature value is $1-2$ orders of magnitude higher than the red horizontal line denoting the assumed intrinsic brightness temperature, $T_{\text{eq}}$. Under the assumption of the intrinsic brightness temperature being close to the equipartition value, we can interpret these data as the result of Doppler boosting present at every epoch. The average brightness temperature is $1.87 \cdot 10^{12}$\,K, and the average Doppler boosting factor is $37$. The highest $T_\mathrm{b, VLBI}$ value is actually a lower limit, since the fitted core size was smaller than the smallest resolvable angular size of the array in that particular epoch. Thus only an upper limit could be given for the size of the core component. 

\subsection{Spectral index}

We calculated the spectral index of the core, $\alpha$, using the convention $S_\mathrm{core}\propto \nu^\alpha$. For the estimation, we selected the 1996 measurements, since in this year, all measurements at 3 different frequencies were conducted close in time, within a few weeks. 
 
The fitted power-law function has an exponent of $\alpha=- 0.29 \pm 0.13$. Alternatively, we can also calculate a two-point spectral index, $\alpha_\mathrm{CX}$, using the simultaneous $8.3$ and $4.9$\,GHz measurements on 2016 July 17. The value obtained is $\alpha_\mathrm{CX}=0.0\pm0.1$.

Both of these spectral indices indicate a flat radio spectrum, which is in line with the general expectations that the cores of blazars have either flat ($\alpha \approx 0$) or inverted ($\alpha \gtrsim 0.5$) spectra, while the jet typically has a steep ($\alpha \lesssim - 0.7$) spectrum \cite{spectral_mojave}.

\section{Discussion}

\begin{figure}
\includegraphics[width=12.5 cm]{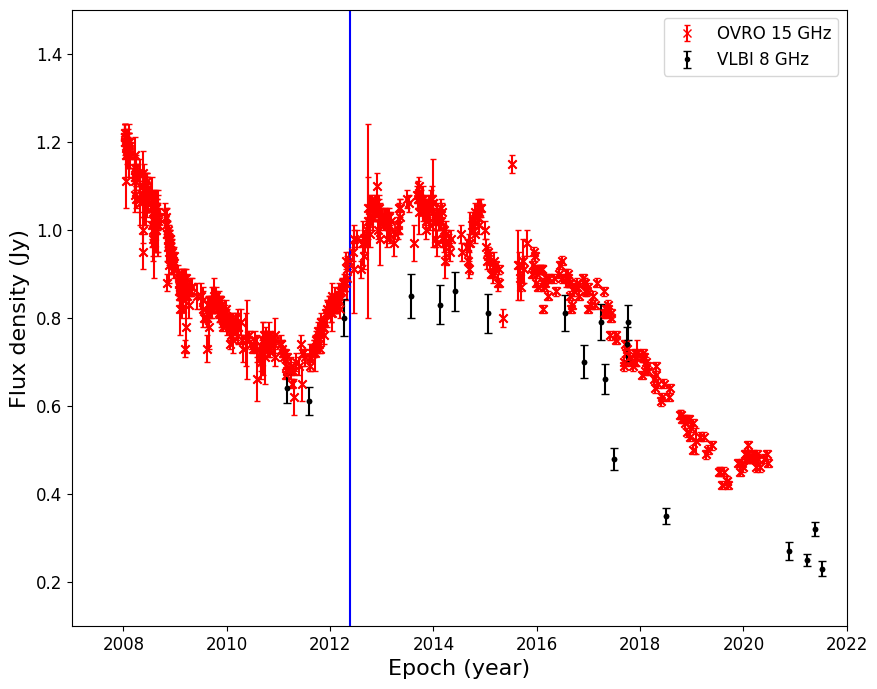}
\caption{The radio light curve of CTD\,74 at $15$\,GHz as measured by the OVRO monitoring program (red crosses, \cite{Hovatta_2021}) along with the flux densities of the core component obtained from the $8$-GHz VLBI measurements (black squares). The time of the neutrino event EHEA2012-05-23 is indicated by the blue vertical line. \label{fig:Hovatta}}
\end{figure}

While CTD\,74 shows brightness temperature values significantly exceeding the equipartition limit, no apparent motion is shown by the jet features  even in the highest-resolution VLBI data at $8.3$\,GHz, therefore superluminal motion could not be derived in its jet. 
At the redshift of the object, a maximum proper motion of $\sim0.1\textrm{\,mas\,yr}^{-1}$ would already imply apparent superluminal motion of a jet feature. Assuming a moderate Lorentz factor of $10$, this would give a critical inclination angle of $1/\Gamma \sim 6^\circ$.
However, the core--jet component separations in the analyzed VLBI data do not allow for such proper motion value (see Figure~\ref{fig:X_sep}). The undetectable proper motions might be attributed to the very small inclination angle of the jet to the line of sight, rendering the movement of the components projected on the sky undetectable. On the other hand, it was shown using probabilistic arguments that all of the BL Lac objects with slow moving components cannot have jets with low Lorentz factors and small viewing angles. It was suggested that in those sources, the pattern speed may not be equal to the speed of the jet beam \cite{cohen2007}. Nevertheless, as long as the proper motion of the jet components cannot be derived, the inclination angle of the jet cannot be safely deduced from the brightness temperature measurements only.

CTD\,74 is part of the sample of blazars monitored at $15$\,GHz with the Owens Valley Radio Observatory (OVRO) as part of the OVRO $40$-m telescope flux density monitoring program \cite{Richards}. For a large number of the monitored sources, the variability brightness temperatures were calculated \cite{Liodakis2018}. For CTD\,74, the obtained value is $\log_{10}\left({\frac{T_\mathrm{b,var}}{1 \mathrm{K}}}\right)=15.2_{-0.42}^{+0.54}$, and the inferred variability Doppler factor is $\delta_\mathrm{var}=56.8^{+26.95}_{-20.26}$. This is the highest Doppler factor among $20$ neutrino-emitter candidate blazars studied using their OVRO $15$-GHz variability light curves \cite{Hovatta_2021}. Within the uncertainties, this value is consistent with the average Doppler factor, $\delta_\mathrm{avg} \approx 37$, obtained from the $8$-GHz VLBI observations. The slightly lower VLBI value may be because the $8$-GHz core can be further resolved at higher frequencies, thus the region responsible for the variability measured at $15$\,GHz (the $15$-GHz core) is more compact. This could be checked with higher-frequency VLBI imaging observations.

The core flux density measured at $8$\,GHz showed significant variability in the analysed VLBI observations (Figure~\ref{fig:flux+int}). At the time and until about $4$\,yr after the high-energy neutrino event, EHEA2012-05-23, the flux density remained in an elevated state, above $0.8$\,Jy. Additionally, CTD\,74 showed significantly higher $T_\mathrm{b,VLBI}$ $14$\,months after the neutrino event. 
The more densely time-sampled long-term OVRO light curve of CTD\,74 shows that the radio brightening seen in the flux density of the $8$-GHz VLBI-detected core began before the neutrino event, in 2011 (\cite{Hovatta_2021}, Figure~\ref{fig:Hovatta}), which time interval was not sampled by VLBI observations. The neutrino event occurred during a relatively steeply rising, $\sim1.5$-yr long section of the $15$-GHz total flux density light curve. According to the OVRO light curve, the initial maximum in late 2012 was followed by at least two smaller flares. After 2016, the flux density gradually decreased and reached its minimum value according to the published OVRO light curve in late 2019. Similar fading can be traced in the $8$-GHz flux density of the VLBI-detected core component. More than $80\%$ of the $8$-GHz VLBI data were obtained after 2010. The OVRO light curve suggests that there was a potentially even brighter flare before 2008, during the time when no $8$-GHz VLBI observation was conducted of CTD\,74, highlighting the importance of the OVRO monitoring program.
 
In the standard shock-in-jet models, it is expected that the radio flares are delayed by a few tens to a few hundreds of days from the higher to the lower frequencies \cite{hovatta_flare}. The closest peak before our highest estimated $T_\mathrm{b,VLBI}$ value (2013 Jul 24) in the OVRO light curve is in late November 2012. However, due to the sparse VLBI observations, we cannot securely connect the $8$-GHz brightening in the VLBI core to any specific feature in the $15$-GHz OVRO light curve. 

An increase in the $10$-GHz radio emission close in time to the high-energy neutrino events was reported in a sample of VLBI-detected blazars selected as potential counterparts due to their positions with respect to the high-energy neutrino events \cite{Plavin}. PKS\,1502$+$106 was identified as showing the highest temporal correlation between the radio flare and the high-energy neutrino event. The neutrino event associated with PKS\,1502$+$106 was detected on 2019 Jul 30 and took place towards the end of a $4$-yr long radio flare seen at various radio frequencies between 2016 and 2020 \cite{plavinsources}. In other two of the reported five high-probability high-energy neutrino candidate sources \cite{plavinsources}, PKS\,1741$-$038 and PKS\,0735$+$178, the corresponding neutrino events coincided with the rising part of major radio flares. Additional flare--neutrino coincidences include most notably TXS\,0506$+$056, as well as PKS\,0215$+$015 \cite{2023arXiv230804311E} and PKS\,1424$-$418. In the latter case, an increase in $\gamma$-ray, X-ray, and optical emission has been reported in temporal and positional coincidence with a PeV energy neutrino event \cite{2016NatPh..12..807K}.
In the context of radio flare--neutrino coincidences, CTD\,74 showed similar behaviour.
However, enhanced flux density variability (across the entire electromagnetic spectrum) is one of the major characteristics of blazars. Therefore a clear connection between the neutrino event and the observed flux density variability in the radio cannot be established for CTD\,74 from the available data. 

High-energy neutrinos can be produced in blazar jets via, e.g., the proton--photon process (see \cite{Plavin_neutrino} and references therein). The ingredients of that, the high-energy photons and protons are principally available in the jets, the former via the synchrotron self-Compton mechanism, while the latter can be accelerated in standing shocks \cite{Plavin_neutrino}. In CTD\,74, all detected VLBI features seem to be stationary (Figure~\ref{fig:X_sep}) which can be explained as the jet having a very small inclination angle to line of sight, but also one (or more) could indeed be stationary, a standing shock formed in the jet.

In the proton--photon formation channel of high-energy neutrinos, photons at GeV energies can also be created in an electromagnetic cascade. However, these photons could hardly be detected due to the expected high optical depth for $\gamma$-rays at this region of the jet \cite{Plavin_neutrino}. Thus, a high-energy neutrino candidate source does not necessarily have to be a $\gamma$-ray loud blazar. CTD\,74 does not appear in the most recent \textit{Fermi}-LAT catalog of $\gamma$-ray-detected extragalactic sources \cite{Abdollahi_2020}.

\section{Summary}

We analyzed a series of VLBI measurements conducted between 1996 and 2021 of CTD\,74, a blazar that was positionally linked to a high-energy neutrino event detected by IceCube. We reconstructed the brightness distribution of the source and subsequently fitted Gaussian model components to the calibrated visibilities. We were able to identify a couple of jet components both at $2$ and $8$\,GHz in a short northwest-directed jet. During the $\sim25$\,yr spanned by the VLBI observations, we did not detect significant motion in any of the jet components. 

The $8$-GHz brightness temperature of the core exceeded the equipartition brightness temperature value in every epoch, implying Doppler boosting of the emission, and Doppler factors of several tens.

The $8$\,GHz flux density of the core component revealed a brightening at around the end of 2011. This several year-long flare can be seen in the densely-sampled $15$-GHz total flux density light curve from the OVRO monitoring program as well. The neutrino event took place during the rising phase of this radio brightening, preceding the maximum by about half a year. 

We found that CTD\,74, which was positionally associated \cite{Plavin} with a high-energy neutrino event in 2012, shows several properties that are in line with other blazars listed as possible neutrino sources. These are the high Doppler factor, the presence of stationary features in the jet, and enhanced flux density variability around the time of the neutrino event. However, since blazars generally have Doppler-boosted jet emission and are variable, a comprehensive theoretical framework of the neutrino emission from blazars, with observationally testable predictions would be essential for reaching firmer conclusion in cases like CTD\,74. Moreover, rapid-response quasi-contemporaneous VLBI imaging of candidate high-energy neutrino source blazars, triggered by new neutrino detections, would help reveal if there is clear connection between the neutrino emission and the changes in the radio structure. Obviously, continuing high-cadence flux density monitoring of a large sample of potential neutrino-emitting blazars is needed for associating the neutrino events to outburst or any other light curve feature.       

\vspace{6pt} 

\authorcontributions{Conceptualization and investigation, K.É.G; formal analysis, J.K. and K.É.G.; writing—original draft preparation, J.K. and K.É.G.; supervision, K.É.G.; writing—review and editing, K.É.G., S.F. and E.K. All authors have read and agreed to the published version of the manuscript.
}

\funding{This research was funded by the Hungarian National Research, Development and Innovation Office (NKFIH), grant number OTKA K134213. This project has received funding from the HUN-REN Hungarian Research Network.}

\dataavailability{The calibrated VLBI data are available from the Astrogeo archive (\url{http://astrogeo.org/} and \url{https://astrogeo.smce.nasa.gov/vlbi_images/}) or from the corresponding author upon reasonable request.}

\acknowledgments{We thank the reviewers for taking the time and effort necessary to review the manuscript. We appreciate all valuable comments and suggestions, which helped us improve the quality of the manuscript.
J.K. gratefully acknowledges the support given by the MTA-ELTE Lend\"ulet ``Momentum'' Milky Way Research Group and its PI, Szabolcs M\'esz\'aros through a student grant. We acknowledge the use of archival calibrated VLBI data from the Astrogeo Center database maintained by Leonid Petrov.}

\conflictsofinterest{The authors declare no conflict of interest.} 

\abbreviations{Abbreviations}{
The following abbreviations are used in this manuscript:\\

\noindent 
\begin{tabular}{@{}ll}
AGN & active galactic nuclei\\
EVN & European VLBI Network\\
FWHM & full width at half maximum\\
ICRF & International Celestial Reference Frame\\
NRAO & National Radio Astronomy Observatory\\
OVRO & Owens Valley Radio Observatory\\
SNR & signal-to-noise ratio\\
VLBA & Very Long Baseline Array\\
VLBI & Very Long Baseline Interferometry\\
\end{tabular}
}

\newpage

\appendixtitles{no} 
\appendixstart
\appendix
\section[\appendixname~\thesection]{}

In the Tables \ref{8ghz}, \ref{2ghz}, and \ref{5ghz}, we list the parameters of Gaussian components fitted to the calibrated VLBI visibility data of CTD\,74. 

\begin{longtable}[c]{lllllll}
\caption{Parameters of the components fitted to the $8$-GHz observations. In the first and second columns, the observational epoch and the name of the components are given. In the following columns, $S$ denotes the flux density of the model components, $P$ indicates their separation from the core, $W_1$ and $W_2$ are the FWHM sizes of major and minor axes of the Gaussian components. (In the case of circular components, only $W_1$ is given.) In the last column, $\Phi$ denotes the position angle of the major axis for elliptical components, measured from north through east.}
\label{8ghz}\\
\hline
\textbf{Epoch} & \textbf{\#} & $\mathbf{S}$ \textbf{(Jy)} & $\mathbf{P}$ \textbf{(mas)} & $\mathbf{W_1}$ \textbf{(mas)} & $\mathbf{W_2}$ \textbf{(mas)} & $\mathbf{\Phi}$ $(^{\circ})$ \\ \hline
\endfirsthead
\multicolumn{7}{c}%
{{\bfseries Table \thetable\ continued from previous page}} \\
\hline
\textbf{Epoch} & \textbf{\#} & $\mathbf{S}$ \textbf{(Jy)} & $\mathbf{P}$ \textbf{(mas)} & $\mathbf{W_1}$ \textbf{(mas)} & $\mathbf{W_2}$ \textbf{(mas)} & $\mathbf{\Phi}$ $(^{\circ})$ \\ \hline
\endhead
\hline
\endfoot
\endlastfoot
\multirow{2}{*}{1996.05.15.} & X0 & $0.82 \pm 0.04$ & $0$ & $0.56 \pm 0.01$ & $0.19 \pm 0.01$ & $-55 \pm 5$ \\
 & X1 & $0.052 \pm 0.004$ & $1.77 \pm 0.29$ & $2.32 \pm 0.03$ &  &  \\ \hline
\multirow{3}{*}{1997.01.10.} & X0 & $0.57 \pm 0.03$ & $0$ & $0.63 \pm 0.01$ & $0.222 \pm 0.004$ & $-60 \pm 5$ \\
 & X1 & $0.020 \pm 0.001$ & $1.85 \pm 0.29$ & $1.50 \pm 0.02$ &  &  \\
 & X2 & $0.01 \pm 0.01$ & $4.43 \pm 0.26$ & $2.99 \pm 1.07$ &  &  \\ \hline
\multirow{3}{*}{2000.05.22.} & X0 & $0.37 \pm 0.02$ & $0$ & $0.32 \pm 0.01$ & $0.150 \pm 0.004$ & $-50 \pm 5$ \\
 & J1 & $0.21 \pm 0.01$ & $0.60 \pm 0.18$ & $0.58 \pm 0.01$ &  &  \\
 & J2 & $0.02 \pm 0.01$ & $2.53 \pm 0.19$ & $1.13 \pm 0.24$ &  &  \\ \hline
\multirow{3}{*}{2002.09.25.} & X0 & $0.42 \pm 0.02$ & $0$ & $0.226 \pm 0.004$ &  &  \\
 & J1 & $0.15 \pm 0.01$ & $0.56 \pm 0.19$ & $0.338 \pm 0.004$ &  &  \\
 & J1.5 & $0.10 \pm 0.01$ & $1.20 \pm 0.21$ & $1.73 \pm 0.01$ &  &  \\ \hline
\multirow{3}{*}{2011.02.27.} & X0 & $0.64 \pm 0.03$ & $0$ & $0.34 \pm 0.01$ &  &  \\
 & J1 & $0.17 \pm 0.01$ & $0.67 \pm 0.35$ & $0.403 \pm 0.003$ &  &  \\
 & J2 & $0.051 \pm 0.003$ & $2.47 \pm 0.33$ & $2.11 \pm 0.05$ &  &  \\ \hline
\multirow{4}{*}{2011.08.06.} & X0 & $0.61 \pm 0.03$ & $0$ & $0.211 \pm 0.003$ &  &  \\
 & J1 & $0.21 \pm 0.01$ & $0.52 \pm 0.28$ & $0.349 \pm 0.002$ &  &  \\
 & J1.5 & $0.040 \pm 0.002$ & $1.27 \pm 0.33$ & $0.787 \pm 0.004$ &  &  \\
 & J2 & $0.039 \pm 0.002$ & $2.66 \pm 0.34$ & $2.76 \pm 0.03$ &  &  \\ \hline
\multirow{3}{*}{2012.04.10.} & X0 & $0.79 \pm 0.04$ & $0$ & $0.167 \pm 0.003$ &  &  \\
 & J1 & $0.23 \pm 0.01$ & $0.67 \pm 0.32$ & $0.438 \pm 0.002$ &  &  \\
 & J2 & $0.062 \pm 0.003$ & $2.47 \pm 0.32$ & $2.77 \pm 0.02$ &  &  \\ \hline
\multirow{3}{*}{2013.07.24.} & X0 & $0.86 \pm 0.05$ & $0$ & $\le 0.05$ &  &  \\
 & J1 & $0.18 \pm 0.01$ & $0.55 \pm 0.14$ & $0.34 \pm 0.06$ &  &  \\
 & J2 & $0.01 \pm 0.01$ & $2.35 \pm 0.17$ & $0.34 \pm 0.43$ &  &  \\ \hline
\multirow{4}{*}{2014.02.12.} & X0 & $0.83 \pm 0.04$ & $0$ & $0.141 \pm 0.003$ &  &  \\
 & J1 & $0.11 \pm 0.01$ & $0.47 \pm 0.17$ & $0.305 \pm 0.002$ &  &  \\
 & J1.5 & $0.036 \pm 0.002$ & $1.06 \pm 0.18$ & $0.47 \pm 0.01$ &  &  \\
 & J2 & $0.04 \pm 0.01$ & $2.61 \pm 0.19$ & $1.83 \pm 0.15$ &  &  \\ \hline
\multirow{3}{*}{2014.05.31.} & X0 & $0.86 \pm 0.04$ & $0$ & $0.290 \pm 0.002$ &  &  \\
 & J1 & $0.12 \pm 0.01$ & $0.70 \pm 0.35$ & $0.537 \pm 0.002$ &  &  \\
 & J2 & $0.047 \pm 0.003$ & $2.39 \pm 0.34$ & $2.23 \pm 0.02$ &  &  \\ \hline
\multirow{3}{*}{2015.01.23.} & X0 & $0.81 \pm 0.04$ & $0$ & $0.241 \pm 0.005$ &  &  \\
 & J1 & $0.17 \pm 0.01$ & $0.69 \pm 0.29$ & $0.611 \pm 0.004$ &  &  \\
 & J2 & $0.02 \pm 0.01$ & $2.52 \pm 0.32$ & $0.76 \pm 0.26$ &  &  \\ \hline
\multirow{3}{*}{2016.07.17.} & X0 & $0.81 \pm 0.04$ & $0$ & $0.252 \pm 0.002$ &  &  \\
 & J1 & $0.10 \pm 0.01$ & $0.71 \pm 0.36$ & $0.624 \pm 0.003$ &  &  \\
 & J2 & $0.034 \pm 0.004$ & $3.05 \pm 0.42$ & $2.05 \pm 0.19$ &  &  \\ \hline 
\multirow{3}{*}{2016.11.30.} & X0 & $0.70 \pm 0.04$ & $0$ & $0.291 \pm 0.004$ &  &  \\
 & J1 & $0.06 \pm 0.01$ & $0.56 \pm 0.18$ & $1.06 \pm 0.03$ &  &  \\
 & J2 & $0.03 \pm 0.09$ & $1.70 \pm 0.18$ & $0.05 \pm 0.12$ &  &  \\ \hline
 \pagebreak
\multirow{3}{*}{2017.03.27.} & X0 & $0.79 \pm 0.04$ & $0$ & $0.241 \pm 0.003$ &  &  \\
 & J1 & $0.084 \pm 0.004$ & $0.90 \pm 0.38$ & $0.440 \pm 0.004$ &  &  \\
 & J2 & $0.031 \pm 0.003$ & $2.64 \pm 0.30$ & $3.15 \pm 0.04$ &  &  \\ \hline
\multirow{3}{*}{2017.04.28.} & X0 & $0.66 \pm 0.01$ & $0$ & $0.233 \pm 0.002$ &  &  \\
 & J1 & $0.17 \pm 0.01$ & $0.46 \pm 0.34$ & $0.33 \pm 0.01$ &  &  \\
 & J2 & $0.048 \pm 0.003$ & $2.11 \pm 0.32$ & $2.01 \pm 0.02$ &  &  \\ \hline
\multirow{3}{*}{2017.06.28.} & X0 & $0.48 \pm 0.03$ & $0$ & $0.337 \pm 0.004$ &  &  \\
 & J1 & $0.29 \pm 0.01$ & $0.62 \pm 0.14$ & $0.11 \pm 0.01$ &  &  \\
 & J2 & $0.06 \pm 0.02$ & $2.27 \pm 0.16$ & $2.11 \pm 0.03$ &  &  \\ \hline
\multirow{3}{*}{2017.09.26.} & X0 & $0.74 \pm 0.04$ & $0$ & $0.342 \pm 0.002$ &  &  \\
 & J1 & $0.077 \pm 0.004$ & $0.92 \pm 0.45$ & $0.554 \pm 0.004$ &  &  \\
 & J2 & $0.048 \pm 0.003$ & $2.50 \pm 0.40$ & $2.38 \pm 0.01$ &  &  \\ \hline
\multirow{3}{*}{2017.10.09.} & X0 & $0.79 \pm 0.04$ & $0$ & $0.471 \pm 0.004$ & $0.23 \pm 0.01$ & $-61 \pm 5$ \\
 & J1.5 & $0.42 \pm 0.02$ & $1.27 \pm 0.40$ & $1.00 \pm 0.03$ &  &  \\
 & J2 & $0.030 \pm 0.002$ & $2.93 \pm 0.33$ & $3.06 \pm 0.01$ &  &  \\ \hline
\multirow{3}{*}{2018.07.05.} & X0 & $0.35 \pm 0.02$ & $0$ & $0.476 \pm 0.004$ &  &  \\
 & J1 & $0.053 \pm 0.003$ & $0.69 \pm 0.27$ & $0.75 \pm 0.01$ &  &  \\
 & J2 & $0.058 \pm 0.004$ & $2.17 \pm 0.27$ & $2.44 \pm 0.08$ &  &  \\ \hline
\multirow{4}{*}{2020.11.18.} & X0 & $0.27 \pm 0.02$ & $0$ & $0.49 \pm 0.03$ &  &  \\
 & J1 & $0.10 \pm 0.01$ & $0.54 \pm 0.22$ & $0.22 \pm 0.02$ &  &  \\
 & J1.5 & $0.01 \pm 0.03$ & $1.34 \pm 0.41$ & $0.05 \pm 0.01$ &  &  \\
 & J2 & $0.039 \pm 0.002$ & $2.40 \pm 0.29$ & $2.21 \pm 0.04$ &  &  \\ \hline
\multirow{4}{*}{2021.03.24.} & X0 & $0.25 \pm 0.01$ & $0$ & $0.24 \pm 0.01$ &  &  \\
 & J1 & $0.15 \pm 0.01$ & $0.52 \pm 0.20$ & $0.29 \pm 0.01$ &  &  \\
 & J1.5 & $0.048 \pm 0.003$ & $1.19 \pm 0.33$ & $1.32 \pm 0.03$ &  &  \\
 & J2 & $0.028 \pm 0.002$ & $2.44 \pm 0.34$ & $2.36 \pm 0.04$ &  &  \\ \hline
\multirow{4}{*}{2021.05.19.} & X0 & $0.32 \pm 0.02$ & $0$ & $0.60 \pm 0.01$ &  &  \\
 & J1 & $0.11 \pm 0.01$ & $0.62 \pm 0.17$ & $0.78 \pm 0.05$ &  &  \\
 & J1.5 & $0.026 \pm 0.002$ & $1.19 \pm 0.23$ & $1.50 \pm 0.02$ &  &  \\
 & J2 & $0.028 \pm 0.002$ & $2.92 \pm 0.26$ & $2.67 \pm 0.04$ &  &  \\ \hline
\multirow{3}{*}{2021.07.07.} & X0 & $0.23 \pm 0.02$ & $0$ & $0.12 \pm 0.01$ &  &  \\
 & J1 & $0.11 \pm 0.01$ & $0.56 \pm 0.14$ & $0.47 \pm 0.02$ &  &  \\
 & J2 & $0.049 \pm 0.003$ & $2.41 \pm 0.15$ & $2.29 \pm 0.06$ &  &  \\ \hline
\end{longtable}

\begin{longtable}[c]{lllllll}
\caption{Parameters of the components fitted to the $2$-GHz observations. Columns are the same as in Table~\ref{8ghz}.}
\label{2ghz}\\
\hline
\textbf{Epoch} & \textbf{\#} & $\mathbf{S}$ \textbf{(Jy)} & $\mathbf{P}$ \textbf{(mas)} & $\mathbf{W_1}$ \textbf{(mas)} & $\mathbf{W_2}$ \textbf{(mas)} & $\mathbf{\Phi}$ $(^{\circ})$ \\ \hline
\endfirsthead
\multicolumn{7}{c}%
{{\bfseries Table \thetable\ continued from previous page}} \\
\hline
\textbf{Epoch} & \textbf{\#} & $\mathbf{S}$ \textbf{(Jy)} & $\mathbf{P}$ \textbf{(mas)} & $\mathbf{W_1}$ \textbf{(mas)} & $\mathbf{W_2}$ \textbf{(mas)} & $\mathbf{\Phi}$ $(^{\circ})$ \\ \hline
\endhead
\hline
\endfoot
\endlastfoot
\multirow{3}{*}{1996.05.15.} & S0 & $1.10 \pm 0.07$ & $0$ & $0.34 \pm 0.02$ &  &  \\
 & S1 & $0.12 \pm 0.01$ & $1.91 \pm 0.77$ & $1.51 \pm 0.01$ &  &  \\
 & S2 & $0.072 \pm 0.004$ & $5.63 \pm 0.89$ & $4.82 \pm 0.05$ &  &  \\ \hline
\multirow{3}{*}{1997.01.10} & S0 & $0.99 \pm 0.20$ & $0$ & $\le 0.27$ &  &  \\
 & S1 & $0.080 \pm 0.004$ & $2.23 \pm 0.88$ & $0.34 \pm 0.01$ &  &  \\
 & S2 & $0.041 \pm 0.002$ & $5.64 \pm 0.96$ & $3.99 \pm 0.05$ &  &  \\ \hline
\multirow{3}{*}{2000.05.22.} & S0 & $0.64 \pm 0.03$ & $0$ & $0.53 \pm 0.01$ &  &  \\
 & S1 & $0.15 \pm 0.01$ & $1.52 \pm 0.64$ & $1.591 \pm 0.004$ &  &  \\
 & S2 & $0.058 \pm 0.004$ & $4.30 \pm 0.55$ & $4.17 \pm 0.08$ &  &  \\ \hline
\multirow{3}{*}{2002.09.25.} & S0 & $0.61 \pm 0.03$ & $0$ & $0.67 \pm 0.01$ &  &  \\
 & S1 & $0.163 \pm 0.009$ & $1.60 \pm 0.73$ & $2.07 \pm 0.02$ &  &  \\
 & S2 & $0.060 \pm 0.004$ & $5.18 \pm 0.60$ & $5.32 \pm 0.19$ &  &  \\ \hline
 \pagebreak
\multirow{2}{*}{2013.07.24.} & S0 & $0.63 \pm 0.03$ & $0$ & $1.01 \pm 0.02$ &  &  \\
 & S1 & $0.15 \pm 0.01$ & $1.48 \pm 0.56$ & $1.51 \pm 0.02$ &  &  \\ \hline
\multirow{3}{*}{2014.02.12.} & S0 & $0.79 \pm 0.05$ & $0$ & $0.92 \pm 0.03$ &  &  \\
 & S1 & $0.17 \pm 0.01$ & $2.40 \pm 0.91$ & $2.62 \pm 0.02$ &  &  \\
 & S2 & $0.012 \pm 0.004$ & $8.63 \pm 0.93$ & $0.44 \pm 0.08$ &  &  \\ \hline
\multirow{3}{*}{2014.05.31.} & S0 & $0.71 \pm 0.01$ & $0$ & $0.74 \pm 0.01$ &  &  \\
 & S1 & $0.150 \pm 0.002$ & $1.94 \pm 1.29$ & $1.81 \pm 0.02$ &  &  \\
 & S2 & $0.046 \pm 0.007$ & $5.46 \pm 1.12$ & $4.60 \pm 0.05$ &  &  \\ \hline
\multirow{2}{*}{2015.01.23.} & S0 & $0.71 \pm 0.04$ & $0$ & $0.83 \pm 0.01$ &  &  \\
 & S1 & $0.20 \pm 0.01$ & $2.01 \pm 1.07$ & $2.66 \pm 0.01$ &  &  \\ \hline
\multirow{2}{*}{2016.11.30.} & S0 & $0.74 \pm 0.04$ & $0$ & $0.89 \pm 0.02$ &  &  \\
 & S1 & $0.10 \pm 0.01$ & $3.07 \pm 0.68$ & $2.74 \pm 0.06$ &  &  \\ \hline
\multirow{2}{*}{2017.03.27.} & S0 & $0.76 \pm 0.04$ & $0$ & $0.67 \pm 0.02$ &  &  \\
 & S1 & $0.11 \pm 0.01$ & $3.01 \pm 1.22$ & $2.90 \pm 0.02$ &  &  \\ \hline
\multirow{3}{*}{2017.04.28.} & S0 & $0.68 \pm 1.33$ & $0$ & $\le 0.37$ &  &  \\
 & S1 & $0.13 \pm 0.01$ & $2.21 \pm 1.18$ & $2.156 \pm 0.002$ &  &  \\
 & S2 & $0.017 \pm 0.002$ & $7.88 \pm 1.07$ & $3.80 \pm 0.15$ &  &  \\ \hline
\multirow{2}{*}{2017.06.28.} & S0 & $0.86 \pm 0.05$ & $0$ & $1.05 \pm 0.02$ &  &  \\
 & S1 & $0.18 \pm 0.01$ & $2.78 \pm 0.59$ & $3.84 \pm 0.07$ &  &  \\ \hline
\multirow{2}{*}{2017.09.26.} & S0 & $0.88 \pm 0.04$ & $0$ & $1.01 \pm 0.01$ &  &  \\
 & S1 & $0.11 \pm 0.01$ & $3.48 \pm 1.47$ & $2.59 \pm 0.02$ &  &  \\ \hline
\multirow{3}{*}{2017.10.09.} & S0 & $0.78 \pm 0.04$ & $0$ & $0.55 \pm 0.01$ &  &  \\
 & S1 & $0.16 \pm 0.01$ & $2.03 \pm 1.14$ & $1.25 \pm 0.01$ &  &  \\
 & S2 & $0.047 \pm 0.002$ & $5.13 \pm 0.96$ & $3.30 \pm 0.02$ &  &  \\ \hline
\multirow{3}{*}{2018.07.05.} & S0 & $0.77 \pm 0.03$ & $0$ & $\le 0.19$ &  &  \\
 & S1 & $0.22 \pm 0.01$ & $1.96 \pm 0.92$ & $1.82 \pm 0.01$ &  &  \\
 & S2 & $0.044 \pm 0.003$ & $5.57 \pm 0.75$ & $5.55 \pm 0.06$ &  &  \\ \hline
\multirow{2}{*}{2020.11.18.} & S0 & $0.49 \pm 0.03$ & $0$ & $1.16 \pm 0.01$ &  &  \\
 & S1 & $0.055 \pm 0.003$ & $2.17 \pm 0.53$ & $1.87 \pm 0.02$ &  &  \\ \hline
\multirow{3}{*}{2021.03.24.} & S0 & $0.34 \pm 0.02$ & $0$ & $1.59 \pm 0.03$ & $0.52 \pm 0.04$ & $-50 \pm 5$ \\
 & S1 & $0.05 \pm 0.01$ & $1.96 \pm 1.11$ & $4.06 \pm 0.03$ &  &  \\
 & S2 & $0.031 \pm 0.002$ & $5.74 \pm 1.28$ & $2.78 \pm 0.02$ &  &  \\ \hline
\multirow{3}{*}{2021.05.19.} & S0 & $0.51 \pm 0.03$ & $0$ & $0.55 \pm 0.03$ &  &  \\
 & S1 & $0.16 \pm 0.01$ & $1.95 \pm 0.83$ & $2.01 \pm 0.01$ &  &  \\
 & S2 & $0.025 \pm 0.002$ & $5.34 \pm 0.65$ & $3.04 \pm 0.09$ &  &  \\ \hline
\multirow{2}{*}{2021.07.07.} & S0 & $0.46 \pm 0.03$ & $0$ & $0.43 \pm 0.02$ &  &  \\
 & S1 & $0.11 \pm 0.01$ & $2.18 \pm 0.51$ & $2.06 \pm 0.07$ &  &  \\ \hline
\end{longtable}

\begin{longtable}[c]{lllllll}
\caption{Parameters of the components fitted to the $5$-GHz observations. Columns are the same as in Table~\ref{8ghz}.}
\label{5ghz}\\
\hline
\textbf{Epoch} & \textbf{\#} & $\mathbf{S}$ \textbf{(Jy)} & $\mathbf{P}$ \textbf{(mas)} & $\mathbf{W_1}$ \textbf{(mas)} & $\mathbf{W_2}$ \textbf{(mas)} & $\mathbf{\Phi}$ $(^{\circ})$ \\ \hline
\endfirsthead
\multicolumn{7}{c}%
{{\bfseries Table \thetable\ continued from previous page}} \\
\hline
\textbf{Epoch} & \textbf{\#} & $\mathbf{S}$ \textbf{(Jy)} & $\mathbf{P}$ \textbf{(mas)} & $\mathbf{W_1}$ \textbf{(mas)} & $\mathbf{W_2}$ \textbf{(mas)} & $\mathbf{\Phi}$ $(^{\circ})$ \\ \hline
\endhead
\hline
\endfoot
\endlastfoot
\multirow{3}{*}{1996.06.05.} & C0 & $0.83 \pm 0.04$ & $0$ & $\le 0.16$ &  &  \\
 & C1 & $0.24 \pm 0.01$ & $0.58 \pm 1.00$ & $0.327 \pm 0.007$ &  &  \\
 & C2 & $0.06 \pm 0.01$ & $3.20 \pm 1.01$ & $2.74 \pm 0.23$ &  &  \\ \hline
\multirow{4}{*}{2006.02.09.} & C0 & $0.65 \pm 0.15$ & $0$ & $0.290 \pm 0.003$ &  &  \\
 & C1 & $0.13 \pm 0.01$ & $1.00 \pm 0.98$ & $0.373 \pm 0.004$ &  &  \\
 & C2 & $0.051 \pm 0.002$ & $2.85 \pm 1.04$ & $3.17 \pm 0.06$ & $0.49 \pm 0.06$ & $41 \pm 5$ \\
 & C3 & $0.02 \pm 0.002$ & $5.74 \pm 1.89$ & $3.60 \pm 0.33$ &  &  \\ \hline
 \pagebreak
\multirow{4}{*}{2016.07.17.} & C0 & $0.81 \pm 0.15$ & $0$ & $0.352 \pm 0.004$ &  &  \\
 & C1 & $0.135 \pm 0.002$ & $1.04 \pm 0.62$ & $1.20 \pm 0.02$ &  &  \\
 & C2 & $0.0429 \pm 0.0004$ & $3.07 \pm 0.61$ & $2.23 \pm 0.01$ & $0.46 \pm 0.01$ & $-4 \pm 5$ \\
 & C3 & $0.008 \pm 0.002$ & $8.31 \pm 2.89$ & $1.48 \pm 0.34$ &  &  \\ \hline
\end{longtable}

\begin{adjustwidth}{-\extralength}{0cm}

\reftitle{References}

\bibliography{bib}

\PublishersNote{}
\end{adjustwidth}
\end{document}